\documentclass{Rinton-P9x6}

\begin{document}

\title{Hidden sector axions: physics and cosmology\footnote{\rm Talk given at
SUGRA20, the International Conference 20 Years of SUGRA and 
Search for SUSY and Unification, Northeastern University, Boston, USA,
March 2003}}

\author{Hans Peter Nilles}

\address{Physikalisches Institut, Universit\"{a}t Bonn, D-53115 Bonn, Germany\\
E-mail: nilles@th.physik.uni-bonn.de}  

\maketitle

\abstracts{
Embedding hidden sector supergravity models in the framework of
string theory leads to the appearance of axion-like degrees of
freedom. Among them is
the model independent axion of heterotic string theory.
It has a decay constant
of order of the Planck scale and could play the role of
a quintessence field.
In models allowing for the required $\mu$ term in the TeV range, 
the hidden sector dynamics leads to a vacuum energy of $(0.003\ {\rm eV})^4$
via a multiple see-saw effect.
A solution to the strong CP-problem is provided by an additional
hidden sector pseudoscalar with properties 
that make it an
acceptable candidate for cold dark matter of the universe.}

\renewcommand{\bea}{\begin{eqnarray}}
\renewcommand{\eea}{\end{eqnarray}}
\def\beq{\begin{equation}}
\def\eeq{\end{equation}}

\def\one{\bf 1}
\def\two{\bf 2}
\def\five{\bf 5}
\def\ten{\bf 10}
\def\tenb{\overline{\bf 10}}
\def\fiveb{\overline{\bf 5}}
\def\threeb{{\bf\overline{3}}}
\def\three{{\bf 3}}
\def\fb{{\overline{F}\,}}
\def\hb{{\overline{h}}}
\def\Hb{{\overline{H}\,}}

\def\slash#1{#1\!\!\!\!\!\!/}
\def\hf{\frac12}

\def\A{{\cal A}}
\def\Q{{\cal Q}}

\def\p{\partial}

\newcommand{\debug}{\emph{!!! CHECK !!!}}

\newcommand{\dd}{\mathrm{d}\,}
\newcommand{\Tr}{\mathrm{Tr}}
\newcommand{\drep}[2]{(\mathbf{#1},\mathbf{#2})}


\def\de{$(0.003\ {\rm eV})^4$}

Supersymmetry might play an important role
in stabilizing the weak scale (of order TeV) against
uncontrolled radiative corrections. Therefore the mechanism of
supersymmetry breakdown is one of the major problems in model
building. It soon became clear that supersymmetric extensions of 
the standard model require supersymmetry breakdown in a hidden 
sector. In its simplest and compelling form, hidden and observable
sector are coupled extremely weakly via interactions of
gravitational strength. The original scheme\cite{Nilles} incorporated 
hidden sector supersymmetry breakdown via gaugino 
condensation\cite{Nilles:1983xx,Ferrara:1983qs}. Susy breakdown by the Polonyi 
mechanism\cite{Polonyi:1977pj} was subsequently discussed 
in\cite{Chamseddine:1982jx,Barbieri:1982eh,Nilles:1983dy}. 
The weak scale (represented by the gravitino mass $m_{3/2}$) 
and the Planck scale $M_{\rm Pl}$ are connected via
a see-saw mechanism
\begin{equation}
m_{3/2}\sim {\frac{M_{\rm SUSY}^2}{M_{\rm Pl}}},
\end{equation}
where $M_{\rm SUSY}\sim 10^{11}$ GeV is the source of 
spontaneous supersymmetry breakdown.

This scheme has a beautiful embedding in the framework of the
$E_8\times E_8$ heterotic string\cite{heterotic}. The interplay of the
3-index antisymmetric tensor field strength and the gaugino condensate
in the hidden $E_8$ 
sector\cite{Derendinger:1985kk,Derendinger:1986cv,Dine:1985rz} 
allows the breakdown of supersymmetry.
In the heterotic M-theory of Horava and 
Witten\cite{Horava:1996qa} the mechanism
persists and the hidden sector obtains a geometrical 
interpretation\cite{Horava:1996vs,Nilles:1997cm,Nilles:1998sx}. 
For a review and references see\cite{Nilles:1983ge,Minnesota}.

These higher dimensional string theories contain many more fields 
that might be relevant for the physics at scales far below the
string scale, particularly a set of pseudoscalar fields that
could be candidates for light axions. In the heterotic theory
there appears the so-called model independent axion\cite{witten},
the pseudoscalar partner of the dilaton. This axion might be 
problematic as the corresponding axion decay constant is
expected to be of the order of string and Planck scale,
causing trouble with a non-zero and large contribution
to the vacuum energy density of the universe\cite{preskill}.
Other axions might be useful as a solution to the strong
CP-problem and/or candidates for cold dark matter in the universe.

Meanwhile
the Type 1a supernova observation of nonzero dark energy
\cite{supernova} makes us believe that the vacuum energy of the
universe is nonzero at a value of approximately
$\lambda^4 \sim (0.003\ {\rm eV})^4$.
This has (re)created a lot of interest in
quintessence models\cite{frieman,hquark1,hquark2,models,natural}.
All these models try to account for the  presently observed 
dark energy, but they differ in the prediction 
of future dark energies. 

In the present talk we want to suggest that the model independent 
axion mentioned above could play the role of such a quintessential
particle (quintaxion) that explains the size of dark energy currently
observed\cite{quintaxion}.
 Models with hidden sector gaugino condensation are
shown to contain a (multiple see-saw)
suppression mechanism for the scalar potential
that leads to $\lambda^4 \sim (0.003\ {\rm eV})^4$. 
One of the reasons for this suppression
is related to the mechanism to solve the so-called $\mu$ problem
of the Higgs mass parameter in supersymmetric models.
The large value of the axion decay constant is now responsible for 
the fact that the quintaxion has not yet settled to its minimal
value, thus giving rise to the dark energy observed. The model
considered contains a second (hidden sector) axion, that mixes with 
the model-independent axion. One linear combination of the two
then plays the role of the quintaxion, while the second is
the invisible QCD-axion for the solution of the strong CP-problem
that simultaneously provides a source for cold dark matter.
This mechanism works because of some interesting relations 
between the mass scales of the model, on one hand 
the similarity of the
scale of  supersymmetry breakdown and the scale of the QCD axion,
on the other hand the coincidence of the vacuum energy and the
mass of the QCD axion. The quintaxion has an extremely small mass
of the order $10^{-32} {\rm eV}$ given by $\lambda^2 / M_{\rm Planck}$.

Such ultra-light pseudo-Goldstone
boson have been 
discussed earlier \cite{frieman,hquark1,hquark2}
in different contexts. In Ref.\cite{frieman},
the mass of the boson was related to the
neutrino mass through $m^2_{\nu}/f$. In Ref.\cite{hquark1,hquark2},
the mass coincided with the
almost massless hidden sector quark(s). 
These models 
need the decay constant around $>10^{17}$~GeV so that
the universe has not yet relaxed to the minimum
of the potential\cite{frieman}.   
If one parametrizes this potential as 
\begin{equation}
V[\phi]\sim \lambda^4 U(\xi),\ \ \xi=\frac{\phi}{f},
\end{equation}
the parameter $\omega=p/\rho$ is expressed as $\omega
=(\frac12\dot\phi^2-V)/(\frac12\dot\phi^2+V)$. The
evolution equation of the quintaxion, $\ddot\phi+
3H\dot\phi+\frac{\p V}{\p\phi}=0$, gives rise to a particular
equation of state. We are interested in the state where
$\ddot\phi$ is negiligible, and obtain
\begin{equation}
\omega\simeq\frac{-6f^2+M_P^2|U'|^2}{6f^2
+M_P^2|U'|^2}
\end{equation}
where $M_P=2.44\times 10^{18}$~GeV 
is the reduced Planck mass and $U'=\p U/\p\xi$.
The quintessence condition $\omega<-\frac13$ requires
that $f>\frac{M_P}{\sqrt3}|U'|$. For example, if
$f\simeq 10^{17}$~GeV, the potential of the form
$-\cos\xi$ requires $\xi=[\pi-0.07,\pi+0.07]$ which 
may not be considered as a fine-tuning. For a larger value 
of $f$ this range is even 
increased \footnote{Due to the large value of $f$, the potential 
is extremely flat and the axion is frozen at its initial value
and thus has not moved in the recent past of the cosmological
history}. 
This shows that
natural quintessence requires $f$ near the 
reduced Planck scale, and the mass of the quintaxion
to be around $10^{-32}$~eV. Depending on the specific
properties of the model under consideration such a field
might be detectable through its cosmological effect
of rotating the polarization state of radiation from
distant sources\cite{carroll}.

The models studied in Ref.\cite{hquark1,hquark2} 
rely on standard axion physics \cite{axionrev}
which we are going to present here for completeness. 
The axionlike 
boson $a_q$ generates a tiny potential. 
In QCD, we know that if there exists a very light up 
quark $u$ then the instanton induced $\theta$
dependent free energy has the form
\begin{equation}\label{mu}
-m_u\Lambda_{QCD}^3\cos\bar\theta\simeq -m_\pi^2 f_\pi^2
\cos\bar\theta
\end{equation}
where $\bar\theta$ and $\Lambda_{QCD}$ are the QCD vacuum
angle and the QCD scale. By 
making $m_u$ small, one can shrink the instanton induced
potential. In Refs.~\cite{hquark1}, this fact was
observed but not applied to a specific model.
In these models with ultralight pseudo Goldstone bosons, it was assumed that
the cosmological constant problem(CCP)\cite{veltman,weinberg}
is solved by some as yet not understood mechanism 
such as the self-tuning solutions\cite{kkl}
or as a consequence of a symmetry \footnote{In the present paper
we adopt the same attitude towards the solution of the CCP.}. 
Because $a_q$ is a pseudo-Goldstone boson,
the difference between the maximum and minimum points of the $a_q$
potential is 2 in units of the explicit breaking scale(of order \de)
of the global symmetry. The solution of the CCP is expected to
be achieved at an extremum point such that equations of motion
determine the vanishing cosmological 
constant.\footnote{Here, we assume that the zero cosmological
constant is reached from above, i.e. in a de Sitter space. Recently,
it has been argued that it is reached from below, i.e. in an anti de 
Sitter space\cite{linde}. In this case also, our argument applies.}     

In Ref.~\cite{hquark1,hquark2}, it was attempted to interpret
a model-dependent axion as the ultra light pseudo Goldstone boson.
In this talk, however, we attempt to interpret the {\it 
model-independent axion(MI-axion)}\cite{witten} 
in superstring models as the quintaxion candidate.

The MI-axion $a_{MI}$ is the pseudoscalar field present in
the two form field $B_{MN}\ (M,N=0,1,2,\cdots,9)$: $\p_\mu a_{MI}
\sim \epsilon_{\mu\nu\rho\sigma}H^{\nu\rho\sigma}\ 
(\mu,\nu=0,1,2,3)$ where $H$ is the field strength of $B$.
In models with an anomalous $U(1)$ gauge symmetry, this MI-axion
is removed from the low energy spectrum and there is no pseudoscalar
degree for quintessence. On the other hand, if
there does not exist such an anomalous $U(1)$ gauge symmetry then the
MI-axion survives down to low energy. But it was noted that
there would appear a cosmological energy crisis \cite{ck} 
of the MI-axion if it were the QCD axion, 
since the decay constant 
is near the Planck scale. 
However, if the potential for the
MI-axion is made very flat so that the universe has not rolled
down the hill yet, then the energy density explains the presently
observed dark energy. 
So the superstring models without the anomalous $U(1)$ gauge
symmetry belong to the class of models we discuss here.

In the gravity mediated supersymmetry breaking scenario via 
the hidden sector gaugino condensation, the mass of the
hidden sector gaugino is of order TeV. 
The height of the hidden sector instanton 
induced potential depends on this gaugino mass.
Note that the current quark mass $m_u$ appears in the
coefficient of instanton induced potential (\ref{mu}).
This happens because the chiral transformation $u\rightarrow
e^{i\gamma_5\alpha}u$ is equivalent to changing the coefficient
of the anomaly term by $\bar\theta\rightarrow\bar\theta-2\alpha$.
Thus, this symmetry manifests itself through the appearance
of the current quark 
mass in Eq.~(\ref{mu}). Similarly,
with gaugino condensation in the hidden sector, the
hidden sector gaugino
mass appears in the coefficient of the instanton induced
potential and hence can influence the height of the potential
significantly in particular for a large hidden sector gauge group,
as we shall see explicitely in the following.

Suppose that the hidden sector gauge group is $SU(N)_h$ and there
are $n$ pairs almost massless hidden sector quarks and 
anti-quarks, transforming like
the (anti-)fundamental representation of $SU(N)_h$. Then, the
coefficient of the hidden sector instanton induced potential is
\begin{equation}\label{potenti}
\lambda_h^4\equiv m_Q^n m_{\tilde G}^N\Lambda_h^{4-n-N}.
\end{equation}  
where $\Lambda_h\simeq 10^{13}$~GeV is the hidden sector
scale and $m_{\tilde G}$ is the hidden sector gaugino mass.

Let us now discuss some illustrative examples for the
conditions between $m_Q,n$ and $N$ needed to account
for the \de dark energy, assuming $m_{\tilde G}\simeq 1$~TeV,
\begin{equation}\label{numbers}
\left(\frac{m_Q}{\Lambda_h}\right)^n\sim \left\{
\matrix{10^{-68}\ \mbox{  for SU(3)}_h\cr 
        10^{-58}\ \mbox{  for SU(4)}_h\cr 
        10^{-48}\ \mbox{  for SU(5)}_h\cr} \right.
\end{equation}
For $N=4$, we obtain $m_Q\simeq 10^{-45}$~GeV, $10^{-16}$~GeV,
and $10^{-7}$~GeV, respectively, for $n=1,2,$ and 3. 

This shows
that the suppression required can be easily obtained: but it is quite
model dependent. In realistic models, however, there are some 
additional constraints on the parameters that are also relevant
for the height of the instanton induced potential.
One of them concerns the
notorious $\mu$ problem\cite{kimnilles} in supergravity.
Contributions to the $\mu$ term could either come from
the superpotenial \cite{kimnilles} or the K\"ahler 
potential \cite{giudice}. Understanding the small size of
$\mu$ requires the presence of a symmetry.
The Giudice-Masiero
mechanism\cite{giudice} also relies on a symmetry 
since here one has to forbid the $H_1H_2$ term in the 
superpotential ($H_1$ and $H_2$ are the Higgs 
doublet superfields giving masses to down and up type 
quarks, respectively). The Peccei-Quinn symmetry is
probably the most plausible symmetry for this purpose.
It can solve the $\mu$-problem and introduce a very light
axion: a possible candidate for cold dark matter(CDM). 
In hidden sector
supergravity models it was shown that 
\begin{equation}\label{Wmu}
W_\mu=\frac{c}{M_P}QQ^cH_1H_2
\end{equation}
can give a reasonable value of $\mu$.
Here $c$ is a constant of order 1,  
and both $Q$ and $Q^c$ are the left-handed hidden
sector quarks transforming like {\bf $N$} and {\bf 
$\bar{N}$} of $SU(N)_h$. The scalar superpartners of $Q$ and $Q^c$
are required to condense at a scale near $\Lambda_h$
without breaking supersymmetry, and this hidden sector squark
condensation generates the needed $\mu$ term\cite{ckn}. 
The hidden sector quarks are
not required to condense, otherwise supersymmetry is broken at
the hidden sector scale. Gauginos can condense without 
supersymmetry breaking at the hidden sector scale, but will
break supersymmetry through gravity mediation.
Eq.~(\ref{Wmu}) is the key equation for this mechanism as it
fulfills two roles. Its first is the generation of the
$\mu$ term through hidden sector squark condensation.
Its second role is the generation of a mass for the otherwise
massless hidden sector quarks, once the Higgs fields acquire
a nonvanishing vacuum expectation value. 

As we can see from equation~(\ref{potenti})
the relevance of this discussion of the $\mu$ term for the
height of the instanton induced potential becomes evident
once we realize that $W_\mu$ contributes to the masses of
the hidden sector quarks when $H_1$ and $H_2$
develop vacuum expectation values(VEV's). Let us now construct
an explicit model
with a Peccei-Quinn symmetry $U(1)_X$. This symmetry is chosen in 
such a way that the dimension-3 mass term of $Q$ can be forbidden 
and $m_Q$
can be made extremely small.

\begin{center}
Table I. {\it The $U(1)_X$ quantum numbers of relevant fields.
}
\vskip 0.3cm
\begin{tabular}{|c|cc|cc|ccc|}
\hline
 & $Q$ & $Q^c$ & $H_1$ & $H_2$ & $q$ & $u^c$ & $d^c$ \\
\hline
$X$ & $1$ & $1$ & $-1$  & $-1$  & $0$  & $1$  & $1$ \\
\hline
\end{tabular}
\end{center}

\noindent The $U(1)_X$ quantum numbers of the hidden sector
quarks $Q$ and $Q^c$, the Higgs supermultiplets, the ordinary
quark doublets $q$, the up type quarks and down type quarks 
are shown in Table I. Then, the hidden sector quark obtains mass
of order
\begin{equation}
m_Q\simeq
0.64\times 10^{-14}\sin 2\beta\ \mbox{[GeV]}.
\end{equation}
where $\tan\beta\ $=$\ 
\langle H^0_2\rangle/\langle H_1^0\rangle$. 
Therefore, in view of Eq.~(\ref{numbers}) two
hidden sector quarks $Q_1, Q_1^c, Q_2, Q_2^c$ in $SU(4)_h$
can generate a reasonable height for the quintessence 
potential provided that there is no other significant contribution.

The VEV of the squark condensate breaks the global $U(1)$ symmetry
spontaneously \footnote{For completeness we also have to take
into account hidden sectro gaugino condensation that  leads to
a breakdown of a different chiral symmetry.
This extra symmetry, however, is explicitly broken 
by the hidden sector gaugino
mass term $-m_{\tilde G}\tilde G\tilde G$.  
Inclusion of this effect is trivial and
the essential features discussed below are
not changed because the corresponding meson obtains
a huge mass 
at the order of the $\sqrt{m_{\tilde G}\Lambda_h}$.}.
The resulting pseudo-Goldstone boson $a_h$ can be
identified through 
\begin{equation}
\langle \tilde Q\tilde Q^c\rangle \equiv \tilde v^2
\exp\left(i \frac{a_h}{F_h}\right)
\end{equation}
where $\tilde v\sim\Lambda_h$ and $F_h\sim\Lambda_h$.
The K$\ddot{\rm a}$hler potential is expected to
respect the $U(1)_X$ symmetry. Therefore, it does not
introduce an important contribution to 
the potential for $a_h$. The 
superpotential also respects the $U(1)_X$ symmetry and
does not generate a potential for $a_h$, either.

However, the hidden sector $SU(N)_h$ and QCD $SU(3)_c$
instantons break the $U(1)$ chiral symmetry 
explicitely and introduce 
anomalous couplings of $a_h$. Given the quantum numbers
of Table I, we obtain
\begin{equation}\label{ahcoupling}
\frac{a_h}{F_h}\frac{2}{32\pi^2} \left[
nF_h\tilde F_h+6F\tilde F
\right]
\end{equation}
where $n$ is the number of the hidden sector quarks, 
and we considered 3 families of standard model fermions.
In Eq.~(\ref{ahcoupling}), we used the abbreviated
notations for the hidden sector and QCD anomalies,
$$
F_h\tilde F_h\equiv \frac12 \epsilon_{\mu\nu\rho\sigma}
F_h^{\mu\nu}F_h^{\rho\sigma},\ \ F\tilde F\equiv
\frac12 \epsilon_{\mu\nu\rho\sigma}F^{\mu\nu}F^{\rho\sigma}.
$$

The model also opens up the 
opportunity to solve the strong CP problem by a very light
axion. The QCD axion arising from $a_h$ is composite~\cite{comp}
contrary to the axion candidates suggested in Ref.~\cite{axions}. 
We employ the nonlinearly realized
shift symmetry of the MI-axion $a_{MI}$ 
which is present in any superstring model
without an anomalous $U(1)$ gauge symmetry. 
The MI-axion coupling to the anomaly is universal
\begin{equation}\label{MIcoupling}
\frac{a_{MI}}{F_{MI}}\frac{2}{32\pi^2} \left[
F_h\tilde F_h+F\tilde F
\right].
\end{equation}

This leads to a rather economic model for
quintessence and the solution of the strong CP-problem.
We have two axions $a_h$ and $a_{MI}$ both of which couple
to the hidden sector anomaly. To pick up the QCD axion $a$ and
quintessence $a_q$, let us
define 
\begin{eqnarray}\label{defa}
&a_h=-a_q\sin\alpha +a\cos\alpha,\ \ \ a_q=-a_h\sin\alpha+a_{MI}
\cos\alpha \nonumber\\
&a_{MI}=a_q\cos\alpha +a\sin\alpha,\ \ \
a=a_h\cos\alpha+a_{MI}\sin\alpha 
\end{eqnarray}

The instanton effects of Eqs.~(\ref{ahcoupling}) and 
(\ref{MIcoupling}) generate potentials for the pseudo-Goldstone
bosons $a_h$ and $a_{MI}$,\footnote{
The runaway potential of the dilaton $S$ and $
\langle\tilde Q\tilde Q^c\rangle$ 
is expected to be stabilized at zero cosmological 
constant. The potential $V$ here arises from the
imaginary parts of $S$ and $\tilde Q\tilde Q^c$.}
\begin{equation}\label{potential}
V\sim -\lambda_h^4\cos\left(n\frac{a_h}{F_h}+
\frac{a_{MI}}{F_{MI}}\right)-\Lambda_{QCD}^4
\cos\left(6\frac{a_h}{F_h}+\frac{a_{MI}}{F_{MI}}\right).
\end{equation}
where the coefficient $\Lambda_{QCD}^4$ is a  
symbolic
representation of $\frac{Z }{(1+Z)^2}f_\pi^2m_\pi^2$
with $Z={m_u}/{m_d}$.
The $2\times 2$ mass square matrix in the $a_h$ and $a_{MI}$
basis becomes

\begin{equation}
M^2=\left(\matrix{
 \frac{6^2\Lambda_{QCD}^4+n^2\lambda_h^4}{F_h^2},\ 
\ \frac{6\Lambda_{QCD}^4+n\lambda_h^4}{F_hF_{MI}} \cr
\frac{6\Lambda_{QCD}^4+n\lambda_h^4}{F_hF_{MI}},\ \
\frac{\Lambda_{QCD}^4+\lambda_h^4}{F_{MI}^2} \cr
}\right)
\end{equation}
from which the determinant of $M^2$ is obtained as
\begin{equation}\label{detm2}
{\rm Det}\ M^2= (n-6)^2\ \frac{\Lambda_{QCD}^4
\lambda_h^4}{F_h^2F_{MI}^2}.
\end{equation}
For $n=6$ we obtain a flat direction, and hence we assume
$n\ne 6$ to generate a tiny potential. 
The dominant term in Eq.~(\ref{potential}) is, of course, the
QCD term since the hidden sector term is suppressed by the
masses of the hidden sector gauginos and hidden sector quarks. 
Thus, the argument of the QCD cosine term is defined as the
light axion $a$ with mass of order $10^{-5}$~eV (as a candidate
for cold dark matter):
\begin{equation}
\frac{a}{F_a}\simeq \frac{6}{F_h}a_h+\frac{1}{F_{MI}}a_{MI}
\end{equation} 
from which we obtain in the limit $F_{MI}\gg F_h$
\begin{equation}
\sin\alpha=\frac{F_h}{\sqrt{36F_{MI}^2+F_h^2}}\simeq 
\frac{F_h}{6F_{MI}},\ \ 
\cos\alpha=\frac{6F_{MI}}{\sqrt{36F_{MI}^2+F_h^2}}
\end{equation}
and determine the light axion( QCD axion) parameters
\begin{equation}\label{fa}
F_a=\frac{F_hF_{MI}}{\sqrt{36F_{MI}^2+F_h^2}}
\simeq \frac{F_h}{6},\ \ m^2_a\simeq
\left(\frac{6\Lambda_{QCD}^2}{F_h}\right)^2.
\end{equation}
Note that the smaller decay constant ($F_a$)
corresponds to the larger ($\Lambda_{QCD}^4$) explicit symmetry 
breaking scale and the larger decay constant ($F_q$) corresponds to
the smaller ($\lambda_h^4$) explicit symmetry breaking scale. 
From Eqs.~(\ref{detm2}) and (\ref{fa}), we obtain the mass of 
the quintaxion $a_q$
\begin{equation}
m^2_q\simeq\left(\frac{(n-6)\lambda_h^2}{6 F_{MI}}\right)^2.
\end{equation}
The quintaxion decay constant is close to $F_{MI}$
\begin{equation}
F_q\simeq\frac{6}{|6-n|} F_{MI}.
\end{equation}
Since $F_{MI}$ is near the Planck scale\cite{ck}, we
obtain a large axion decay constant near
that scale, as required for quintessence\cite{frieman}. 

In axion models, it is important to know the domain wall
number and the axion coupling to matter fermions. 
On one hand one has to worry about a possible domain wall 
problem\cite{sikivie} in standard big bang cosmology.
However, in inflationary models with the reheating temperature
below $10^9$~GeV required from the gravitino constraint, 
this old domain
wall problem is only of academic interest. 
The model we presented here has the domain wall number one,
as the MI-axion has the domain
wall number one\cite{dwnumber}. 
The axion-matter coupling in our model is the
same as those of the DFSZ \cite{{Zhitnitsky:tq},{Dine:1981rt}}
 model because the symmetry 
$U(1)_X$ assigns the quantum numbers of the DFSZ model
as shown in Table I. 

Invisible axion models that give suitable candidates for cold
dark matter (CDM) of the universe have to answer the question:
$\lq\lq$Why is $F_a$ near the scale
of the CDM axion?" 
Besides being economic, the model presented here gives an 
explanation for this scale problem.
The breaking scale of the 
Peccei-Quinn symmetry 
is the scale of the hidden sector scalar-quark
condensate. The scale for this condensate is at the
intermediate scale as the requirement 
for the appearance of the 100~GeV scale in the observable sector
should arise from gravity
mediation. In addition, the seed for the $\mu$ term is at this scale, 
and this gives the required axion decay constant of the order of
$10^{12}$~GeV.

We thus have constructed a simple scheme that combines a
mechanism for cold dark matter with one for the dark energy
of the universe. The model contains a light CDM axion
(to solve the strong CP problem)
with decay constant $F_a\sim 10^{12}$~GeV
(through hidden sector squark condensation)
and a quintaxion (reponsible for dark energy)
with $F_q\sim 10^{18}$~GeV
(as expected for the MI-axion).
The potential of the
quintaxion is so shallow because of the smallness of the
hidden sector quark masses which in turn is connected
to the generation of the $\mu$ term.
The main formula that is reponsible for the mechanism discussed here
is eq.~(\ref{Wmu}). It gives the suitable value for the $\mu$ term
as a result of the condensation of hidden sector squarks, but it is
also reponsible for the mass of the hidden sector quarks that appear
once the Higgs bosons receive a nontrivial vacuum expectation value.
It is this multiple see-saw mechanism that leads to the small value
of the vacuum energy of $(0.003\ {\rm eV})^4$
and the extremely small value of the
quintaxion mass of $10^{-32}$~eV.

\vskip 0.4cm
{\bf Acknowledgements}\\[1ex]
\noindent This work was partly supported by 
the European Community's Human Potential
Programme under contracts HPRN--CT--2000--00131 Quantum Spacetime,
HPRN--CT--2000--00148 Physics Across the Present Energy Frontier
and HPRN--CT--2000--00152 Supersymmetry and the Early Universe.

%

\end{document}